\documentclass[aps,prd,twocolumn,superscriptaddress,showpacs,showkeys]{revtex4}

\usepackage[dvips]{graphicx}
\usepackage[english]{babel}

\usepackage{amssymb,epsf,epsfig}
\usepackage{amsmath}

\usepackage{graphicx}

\usepackage{slashed}
\usepackage[active]{srcltx}

\newcommand{\beq}[1]{
\begin{equation}\label{#1}}

\newcommand{\eeq}{\end{equation}}

\newcommand{\bea}[1]{
\begin{eqnarray}\label{#1}}

\newcommand{\eea}{\end{eqnarray}}

\newcommand \vev [1] {\langle{#1}\rangle}
\newcommand \VEV [1] {\left\langle{#1}\right\rangle}

\begin{document}

%%%%%%%%%%%%%%%%%%%%%%%%%%%%%%%%%%%%%%%%%%
%Define Title, Author, Address, Preprint#

%\preprint{ \vbox{  \hbox{CALT-68-}}}

\title{
%\flushright  DESY-15-108\\[3mm]
Evolution equation for the higher-twist B-meson distribution amplitudes}

%\begin{titlepage}

\date{\today}

\author{V.M.~Braun}
\affiliation{Institut f\"ur Theoretische Physik,  Universit\"at
   Regensburg, D-93040 Regensburg, Germany}
\author{A.N.~Manashov}
\affiliation{Institut f\"ur Theoretische Physik, Universit\"at Hamburg, D-22761 Hamburg, Germany}
\affiliation{Institut f\"ur Theoretische Physik, Universit\"at
   Regensburg, D-93040 Regensburg, Germany}
\affiliation{Department of Theoretical Physics,  St.-Petersburg State University, 199034, St.-Petersburg, Russia}
\author{N.~Offen}
\affiliation{Institut f\"ur Theoretische Physik, Universit\"at
   Regensburg, D-93040 Regensburg, Germany}

\date{\today}

\begin{abstract}
We find that the evolution equation for the three-particle quark-gluon B-meson light-cone distribution amplitude
(DA) of subleading twist is completely integrable in the large $N_c$ limit and can be solved exactly. The lowest
anomalous dimension is separated from the rest, continuous, spectrum by a finite gap. The corresponding
eigenfunction coincides with the contribution of quark-gluon states to the two-particle DA $\phi_-(\omega)$ so that the
evolution equation for the latter is the same as for the leading-twist DA $\phi_+(\omega)$ up to a constant shift in
the anomalous dimension.
Thus  ``genuine'' three-particle states that belong to the continuous spectrum effectively decouple from
$\phi_-(\omega)$ to the leading-order accuracy. In turn, the scale dependence of the full three-particle DA 
turns out to be nontrivial so that the contribution with the lowest anomalous dimension does not become leading
at any scales. The results are illustrated on a simple model that can be used in studies of $1/m_b$ corrections to 
heavy-meson decays in the framework of QCD factorization or light-cone sum rules.
%typical observables to the leading-order accuracy.
%Our results suggest that the study of $1/m_b$ corrections to heavy-meson decays in the framework of QCD
%factorization or light-cone sum rules may require a simpler nonperturbative input than it is usually assumed.
 \end{abstract}

%\preprint{DESY ??-???}


\maketitle

%%%%%%%%%%%%%%%%%%%%%%%%%%%%%%%%%%%%%%%%%%%%%%%%%%

\section{Introduction}
%\noindent
%{\large \bf 1.}~~
B-meson light-cone distribution amplitudes (DAs)
are the main nonperturbative input to the
QCD description of weak decays involving light hadrons in the final state~\cite{Beneke:1999br,Beneke:2000wa}.
In particular the leading-twist DA gives a dominant contribution in the heavy quark expansion and it received considerable
attention already~\cite{Grozin:1996pq,Lange:2003ff,Braun:2003wx,Lee:2005gza,Bell:2013tfa,Braun:2014owa,Feldmann:2014ika}.
Utility of the QCD factorization techniques depends, however, on the possibility to control, or at least estimate,
the corrections suppressed by powers of the $b$-quark mass that involve higher-twist DAs.
This task is attracting increasing attention and in the last years there have been several efforts to combine
light-cone sum rules with the expansion in terms of B-meson DAs~\cite{Khodjamirian:2006st,DeFazio:2007hw,Braun:2012kp,Wang:2015vgv}.
This technique allows one to tame infrared divergences which appear power-suppressed contributions in the purely
perturbative framework and to calculate the so-called soft or end-point nonfactorizable contributions in
terms of the DAs of increasing twist. One of the problems on this way is that higher-twist B-meson DAs involve
contributions of multiparton states and are practically unknown.

In this letter we point out that the structure of subleading twist DAs is simpler as compared to what one may assume from
their general partonic decomposition~\cite{Kawamura:2001jm,Nishikawa:2011qk}.
This structure is revealed by considering the scale dependence of the DAs in
the limit of large number of colors, $N_c\to\infty$, i.e. neglecting the $1/N_c^2$ corrections to the renormalization
group equations. It turns out that the evolution equation for the three-particle DA in this approximation is completely integrable
and can be solved exactly. The lowest anomalous dimension is separated
from the rest, continuous, spectrum by a finite gap. The corresponding eigenfunction defines what can be called the ``asymptotic''
three-particle B-meson DA and has a relatively simple form. Most remarkably, it turns out that
the higher-twist contribution to the two-particle B-meson DA $\phi_-(\omega)$ that is related to the three-particle DA
by QCD equations of motion (EOM), is expressed entirely in terms of this ``asymptotic'' state,
the states that belong to the continuous spectrum do not contribute. As the result the DA  $\phi_-(\omega)$ evolves autonomously and
does not mix with ``genuine'' three-particle contributions. The evolution equation for $\phi_-(\omega)$ is the same as
for the leading-twist DA $\phi_+(\omega)$ up to a constant shift in the anomalous dimension.
%Finally, we suggest a simple model for $\phi_-(\omega)$ that can be used in phenomenology.
Finally, we discuss the evolution of the three-particle DA itself and its asymptotic behavior at small and large quark/gluon
momenta which turns out to be nontrivial. 
This behavior is illustrated on the example of a simple model that can be used in phenomenological applications.

%%%%%%%%%%%%%%%%%%%%%%%%%%%%%%%%%%%%%%%%%%%%%%%%%%

%\vskip5mm

\section{Evolution equations}
%
%\noindent
%{\large \bf 2.}~~
Following the established conventions~\cite{Grozin:1996pq} we define the B-meson DAs as matrix elements of the
renormalized nonlocal operators built of an effective heavy quark field $h_v(0)$, a light (anti)quark and gluons
at a light-like separation:
\begin{align}
i F(\mu) \Phi_+(z,\mu)&= \langle 0| \bar q(nz) \slashed{n}\gamma_5 h_v(0) |\bar B(v)\rangle,
\nonumber\\
i F(\mu) \Phi_-(z,\mu) &= \langle 0| \bar q(nz)\slashed{\bar n} \gamma_5  h_v(0) |\bar B(v)\rangle
 \end{align}
and
\begin{eqnarray}
\lefteqn{-2i F(\mu) \Phi_3(z_1,z_2,\mu) =}
\nonumber\\
&=& \langle 0| \bar q(nz_1) gG_{\mu\nu}(nz_2 ) n^\nu \sigma^{\mu\rho}n_\rho\gamma_5 h_v(0) |\bar B(v)\rangle.
\label{three}
\end{eqnarray}
Here $v_\mu$ is the heavy quark velocity,
$n_\mu$ is the light-like vector, $n^2=0$, such that $n\cdot v=1$,
$\Gamma$ stands for an arbitrary Dirac structure,
$|\bar B(v)\rangle$ is the $\bar B$-meson state,
$\mu$ is the factorization scale and $F(\mu)$ is the B-meson decay constant in the heavy quark effective theory (HQET).
Wilson lines connecting the fields
are not shown for brevity; they are always implied.

The functions $\Phi_+$ and $\Phi_-$ are the leading- and subleading-twist two-particle B-meson DAs~\cite{Beneke:2000wa},
and $\Phi_3$ is the (lowest twist) three-particle DA that is the only one relevant for the
present study. In notations of \cite{Kawamura:2001jm} $\Phi_3 = {\Psi}_A-{\Psi}_V$.
These three DAs are related by an EOM~\cite{Beneke:2000wa,Kawamura:2001jm}
\begin{align}
(1+z \partial_z)\,\Phi_-(z)
= \Phi_+(z)  + 2 \int_0^z\! wdw\,\Phi_3(z, w)\,
\label{EOM}
\end{align}
that can be solved to obtain $\Phi_- $ as a sum of the so-called Wandzura-Wilczek (WW) term expressed
in terms of $\Phi_+$~\cite{Beneke:2000wa}, and a certain integral of the quark-gluon DA  $\Phi_3$.
The latter contribution is nontrivial because it involves a function of two variables. We will demonstrate, however,
that this complication is to a large extent illusory as the integral appearing in the EOM essentially
decouples from ``genuine'' quark-gluon correlations. This simplification is exactly analogous to what has been observed
before~\cite{Ali:1991em,Balitsky:1996uh,Braun:2000av,Braun:2001qx} for the structure function $g_2(x,Q^2)$
in polarized deep-inelastic lepton-proton scattering.

%%%%%%%%%%%%%%%%%%%%%%%%%%%%%%%%%%%%%%%%%%%%%%%%%%

%\vskip5mm

%\noindent
%{\large \bf 3.}~~
The following discussion is based on properties of the renormalization group
equations for heavy-light operators under collinear conformal transformations.
 The corresponding generators read
\begin{align}
\label{generators}
  S_+ = z^2 \partial_z + 2j z \,, && S_0 = z\partial_z + j\,, && S_- = -\partial_z\,,
\end{align}
where $j$ is the conformal spin, $j_q=1$ for the light quark and $j_g=3/2$ for the gluon.
The generators  satisfy the standard $SL(2)$ commutation relations
$  [S_+,S_-] = 2 S_0\,,\quad [S_0,S_\pm] = \pm S_\pm\,. $
We distinguish the generators acting on quark and gluon coordinates by the subscript
$S_q$ and $S_g$, respectively.

The starting observation is that both the one-loop renormalization group equations (RGE) for the DAs and the EOM relations
are invariant under special conformal transformations~\cite{Knodlseder:2011gc,Braun:2014owa}.
It is therefore natural to expand the DAs in terms of the  eigenfunctions of
the corresponding generator~\cite{Braun:2014owa}
\begin{align}
Q_s^{(j)}(z)=\frac{e^{-i\pi j}}{z^{2j}} e^{is/z}\,, &&
iS^{(j)}_+ Q_s^{(j)}%(z)
=s Q_s^{(j)}%(z)
\,.
\end{align}
They  form a complete orthonormal set
\begin{align}
&\langle{Q_s^{(j)}|Q_{s'}^{(j)}\rangle}_j={\Gamma(2j)}{s^{1-2j}}\,\delta(s-s')\,,
\\&
\frac1{\Gamma(2j)}\int_0^\infty ds \, s^{2j-1}\,Q_s^{(j)}(z)\,\overline{Q_s^{(j)}(z')} = \frac{e^{-i\pi j}}{(z-\bar z')^{2j}}\,
\notag
\end{align}
 with respect to the $SL(2)$ invariant scalar product~\cite{Gelfand}
\begin{align}
\langle\Phi_1|\Phi_2\rangle_j &=\int_{\mathbb{C}_-} \mathcal{D}_j z\,{\Phi^\ast_1(z)}\,\Phi_2(z)\,,
\label{SP}
\end{align}
where the integration goes over the complex coordinates $z$ in the lower half-plane $\mathbb{C}_-: \text{Im}\,z < 0$
and the integration measure is defined as
$$\mathcal{D}_j z=\frac{2j-1}\pi d^2z\, [i(z-\bar z)]^{2j-2}.$$
Going over from quark/gluon coordinates to the corresponding momenta
\begin{align}
\Phi (z) &= \int_0^\infty \!d\omega\, e^{-i\omega z} \phi(\omega)\,,
\end{align}
can be done easily making use of the following expressions~\cite{Braun:2014owa}:
\begin{align}
\langle{e^{-ikz}|e^{-ik'z}\rangle}_j & ={\Gamma(2j)}\,{k^{1-2j}}\,\delta(k-k')\,,
\notag\\[2mm]
\langle e^{-ikz}|Q_s^{(j)}\rangle_j & = \Gamma(2j)\,(ks)^{1/2-j}\, J_{2j-1}(2\sqrt{ks})\,.
\label{Fourier1}
\end{align}

Staying in coordinate space for the time being, we write the two-particle DAs as
\begin{align}\label{expphipm}
 \Phi_+(z) &= %\int_0^\infty \!ds\, s\,  Q_s^{(1)}(z) \widetilde\phi_+(s,\mu)
-\frac{1}{z^2}  \int_0^\infty \!ds\, s\, e^{is/z} \widetilde\phi_+(s) \,,
% \qquad  \phi_+(s,\mu) = \langle Q_s^{(1)}|\Phi_+\rangle_1\,,
\nonumber\\
 \Phi_-(z) &= %\int_0^\infty \!ds\, Q_s^{(\frac12)}(z) \widetilde\phi_-(s,\mu)
-\frac{i}{z} \int_0^\infty \!ds\, e^{is/z} \widetilde\phi_-(s) \,,
% \qquad  \phi_-(s,\mu) = \langle Q_s^{(\frac12)}|\Phi_-\rangle_{\frac12}\,,
\end{align}
and the three-particle DA
\begin{flalign}
 \Phi_3(z_1,z_2) &=
  \frac{-i}{z_1^2z_2^3}
 \int\limits_0^\infty \!ds \, s^4\! \int\limits_0^1\! \!du\, u \bar u^2\,
e^{is(\frac{u}{z_1}+\frac{\bar u}{z_2})}
\widetilde \phi_3(s,u).&
\notag\\[-3mm]
\label{tildePsi}
\end{flalign}
Here and below $\bar u = 1-u$. Inserting these expressions
in the EOM relation (\ref{EOM}) one derives for the expansion coefficients
\begin{align}
\widetilde\phi_-(s,\mu)  &=  \widetilde\phi_+(s,\mu)
- 2 s^2 \int_0^1 \!du\, u \bar u\, \widetilde \phi_3(s,u,\mu) \,.
\label{tildeEOM}
\end{align}
Invariance under special conformal transformations implies that terms with different values of $s$ cannot
get mixed by the RGE. Thus the leading twist contributions
$\widetilde\phi_+(s,\mu)$ must have autonomous scale dependence:
\begin{align*}
 \Big(\mu\frac{\partial}{\partial\mu}+\beta(g) \frac{\partial}{\partial g}+\frac{\alpha_s}{2\pi}\mathcal{E}_+(s,\mu)\Big)
F(\mu) \, \widetilde \phi_+(s,\mu)=0\,,
\end{align*}
where \cite{Bell:2013tfa,Braun:2014owa}
\begin{align}\label{Eplus}
\mathcal{E}_+(s,\mu) &= 2C_F\left[\ln \big(\mu s \big)-\psi(1)-5/4\right]\,.
%\notag\\
%  & \sigma_q=3/4\,, \qquad \sigma_h=1/2\,,
\end{align}
The RGE for the three-particle DA $\Phi_3$ is more complicated,
\begin{align}
 \Big(\mu\frac{\partial}{\partial\mu}+\beta(g) \frac{\partial}{\partial g}+\frac{\alpha_s}{2\pi}\mathcal{H}\Big)
F(\mu) \, \Phi_3(z_1,z_2,\mu)=0\,,
\label{RGE}
\end{align}
where the ``Hamiltonian'' $\mathcal{H}$ to the one-loop accuracy is given by a sum of two-particle kernels
\begin{align}
  \mathcal{H} = H_{qg} + H_{gh} + H_{qh} \,.
\end{align}
Explicit expressions for the kernels
are known~\cite{Bukhvostov:1985rn,Braun:2009mi,Braun:2014owa,Knodlseder:2011gc}:
\begin{align}
[{H}_{qh}f](z_1) &=-\frac{1}{N_c}\biggl\{\int_0^1\frac{d\alpha}\alpha\Big[f(z_1)-\bar\alpha f(\bar \alpha z_1)\Big]
\notag\\
      &~~+\Big[\ln(i\mu z_1)-\frac54\Big] f(z_1)\biggr\},
\nonumber\\
[{H}_{gh}f](z_2) &= N_c\biggl\{\int_0^1\frac{d\alpha}\alpha\Big[f(z_2)-\bar\alpha^2 f(\bar \alpha z_2)\Big]
\notag\\
      &~~+\Big[\ln(i\mu z_2)-\frac12\Big] f(z_2)\biggr\},
\notag\\
{}[{H}_{qg}\varphi](z_1,z_2)&=
N_c\biggl\{\int_0^1\frac{d\alpha}{\alpha}\Big[2\varphi(z_1,z_2) - \bar\alpha\varphi(z_{12}^\alpha,z_2)
\notag\\
&~~
-\bar\alpha^{2}\varphi(z_1,z_{21}^\alpha)\Big] - \frac34 \varphi(z_1,z_2)\biggr\}
\notag\\
&~~ - \frac2{N_c}\int_0^1 d\alpha\int_{\bar\alpha}^1 d\beta \,\bar\beta\,
\varphi(z_{12}^\alpha,z_{21}^\beta)\,,
\label{Hj}
\end{align}
where
\begin{align}
 z_{12}^\alpha = \bar\alpha z_1 + \alpha z_2\,, && \bar \alpha = 1-\alpha\,.
\end{align}
Note that in difference to \cite{Braun:2014owa,Knodlseder:2011gc,Braun:2009mi} we include the QCD coupling
in the definition of the quark-antiquark-gluon operator, $G_{\mu\nu}\mapsto gG_{\mu\nu}$. This redefinition
affects the constant terms in the kernels.

For our present purposes it is convenient to write these integral operators in terms of the generators
of $SL(2)$ transformations~\cite{Bukhvostov:1985rn,Braun:2014owa}
\begin{align}
H_{qg} & = N_c\Big[\psi\left(J_{qg}\!+\!3/2\right)+\psi\left(J_{qg}\!-\!3/2\right)-2\psi(1)-3/4\Big]
\notag\\
&\quad
+\frac2{N_c}{(-1)^{J_{qg}-3/2}} \frac{\Gamma\left(J_{qg}-3/2\right)}{\Gamma\left(J_{qg}+3/2\right)}\,,
\notag\\
H_{gh} & = N_c \left[\ln \big(i\mu S^+_{g}\big)-\psi(1)-1/2\right]\,,
\notag\\
H_{qh} & = -\frac1 {N_c}\left[\ln \big(i\mu S^+_{q}\big)-\psi(1)-5/4\right]\,,
\end{align}
where $J_{qg}$ is defined in terms of the corresponding quadratic Casimir operator
$ J_{qg}(J_{qg}-1) = (\vec{S}_q+\vec{S}_g)^2 $.
This representation makes manifest that the Hamiltonian $\mathcal{H}$ commutes with the generator of special conformal transformations
\begin{eqnarray}
  \mathbb{Q}_1 &=& i\big(S^+_{q}+S^+_{g}\big)\,,
\label{Q1}
\end{eqnarray}
and therefore the RGE (\ref{RGE}) is ``diagonal'' in $s$.  This symmetry alone is not sufficient, however, to find the solution
since the problem has two degrees of freedom --- the light-cone coordinates of the light quark and the gluon.
It turns out, however, that for the leading contribution for a large number of colors
\begin{align}
  \mathcal{H} = N_c\, \mathbb{H} + {N^{-1}_c}\, \delta\mathbb{H}\,,
\end{align}
there is an additional ``hidden'' symmetry.
Namely, it is possible to construct one more ``conserved charge'', $\mathbb{Q}_2$,
that commutes both with $\mathbb{Q}_1$ and the large-$N_c$ Hamiltonian $\mathbb{H}$:
\begin{align}\label{QHcomm}
{}[\mathbb{Q}_1,\mathbb{Q}_2] = [\mathbb{Q}_1,\mathbb{H}] = [\mathbb{Q}_2,\mathbb{H}] = 0\,.
\end{align}
Having two conserved charges for a problem with two degrees of freedom allows one to diagonalize the Hamiltonian,
i.e. in our case find the multiplicatively renormalizable operators, without the need to solve the RGE equation (\ref{RGE}) explicitly.
This property is known as complete integrability.

The explicit expression for $\mathbb{Q}_2$ can be found using the
formalism of the quantum inverse scattering method (QISM)~\cite{Faddeev:1979gh}:
\begin{align}
 \mathbb{Q}_2 &= \frac94 i S_q^+ \!- iS_g^+\big(S_g^+ S_q^- \!+ S_g^0 S_q^0\big)
\notag\\&\quad
- iS_g^0\big(S_q^0S_g^+\!-S_g^0 S_q^+)\,.
\end{align}
In this approach the charges $\mathbb{Q}_1, \mathbb{Q}_2$ appear in
the expansion of the element $C(u)$ of the monodromy matrix for an open spin chain,
$C(u)\propto u^2\,\mathbb{Q}_1+\mathbb{Q}_2.$
The commutation relation $[C(u),\mathbb{H}]=0$ can be verified by a direct calculation
using the coordinate-space representation for the kernels as given in Eq.~(\ref{Hj}).
%can be found in~\cite{Braun:2009mi,Braun:2014owa,Knodlseder:2011gc},
or, more elegantly, with the help of the QISM techniques. This derivation will be given elsewhere~\cite{BM}.

The ``conserved charges'' $\mathbb{Q}_1$, $\mathbb{Q}_2$ and the ``Hamiltonian'' $\mathbb{H}$
are self-adjoint operators with respect to  the  $SL(2)$ scalar product (\ref{SP}):
\begin{align*}%\label{sc}
\VEV{\Psi|\Phi}%_{1,\frac32}
= \int_{\mathbb{C}_-} \mathcal{D}_1 z_1\int_{\mathbb{C}_-}\mathcal{D}_{\frac32}z_2\,{\Psi^\ast(z_1,z_2)}\Phi(z_1.z_2)\,.
\end{align*}
It follows that they have real eigenvalues and can be diagonalized  simultaneously:
\begin{eqnarray}
 \mathbb{Q}_1 Y_{s,x}(z_1,z_2) &=& s\, Y_{s,x}(z_1,z_2)\,, %\quad\quad\qquad s > 0
\nonumber\\
 \mathbb{Q}_2 Y_{s,x}(z_1,z_2) &=&-s x^2\,Y_{s,x}(z_1,z_2)\,, %\qquad x^2 \in \mathbb{R}
\nonumber\\
 \mathbb{H}\,   Y_{s,x}(z_1,z_2) &=& \mathbb{E}{(s,x)}\, Y_{s,x}(z_1,z_2)\,.
\end{eqnarray}
Note that we write the eigenvalues of $\mathbb{Q}_2$ as a product $-s x^2$ where $s > 0$ is an eigenvalue of
$\mathbb{Q}_1$ and $x^2$ is a real number (but not necessarily positive), $x^2 \in \mathbb{R}$. This structure is
motivated by QISM~\cite{BM}. The eigenfunctions   $Y_{s,x}(z_1,z_2)$ are labeled by two ``quantum numbers'', $s$ and
$x$, and provide the basis of the so-called Sklyanin's representation of Separated Variables~\cite{Sklyanin:1991ss}. They can be found
using the method developed in~\cite{Derkachov:2003qb},
\begin{align}
Y_{s,x}(z_1,z_2) &=\frac{is^2}{z_1^2 z_2^3}\int_0^1 du\,u\bar  u\, e^{is(u/z_1+\bar u/z_2)}\,
\notag\\
&\times\,
{}_2F_1\left(\genfrac{}{}{0pt}{}{-\frac12-ix,-\frac12+ix}{2}\Big|-\frac{u}{\bar u}\right).
\label{Ysx}
\end{align}
The functions $Y_{s,x}$ are symmetric under reflection $x\to -x$.
Since the eigenvalue $x^2$ has to be real,  $x$ can take real or imaginary values.
It is possible to show that for imaginary $x$ there exists only one normalizable solution corresponding
to the particular value $x=i/2$. For this special solution the hypergeometric function disappears and
the eigenfunction becomes very simple:
\begin{align}
Y^{(0)}_{s}(z_1,z_2)=\frac{is^2}{z_1^2 z_2^3}\int_0^1 du\, u\bar  u\,e^{is(u/z_1+\bar u/z_2)}\,.
\label{Y0}
\end{align}
This solution has the lowest energy
\begin{align}
\mathbb{E}_{0}~\equiv~\mathbb{E}(s,x={i}/{2})
= \ln(\mu s)-\psi(1)-1/4\,
\end{align}
and can be interpreted as the ground state of the large-$N_c$ Hamiltonian.
It describes the ``asymptotic'' quark-gluon DA with the lowest
anomalous dimension. The state is normalized as
\begin{align}
\vev{Y^{(0)}_{s}|Y^{(0)}_{s'}}=\delta(s-s')\,.
\end{align}
The eigenfunctions corresponding to real values of $x$ belong to the
continuous spectrum. They are orthogonal to the ground state,
$\langle Y_{s,x}|Y^{(0)}_{s'}\rangle=\ 0$,
and normalized as
\begin{align}
\VEV{Y_{s,x}|Y_{s',x'}}=\delta(s-s')\delta(x-x')\frac{\coth\pi x}{x(x^2+9/4)}\,.
%\qquad \VEV{Y_{s,x}|Y^{(0)}_{s'}}=\ 0\,.
\end{align}
The corresponding eigenvalue (energy) is
\begin{align}
\mathbb{E}(s,x)&= \ln(\mu s)+ \psi\big(3/2+ix\big)+ \psi\big(3/2-ix\big)
\nonumber\\&\quad-3\psi(1)-5/4\,.
\end{align}
The gap between the ground state and the continuous spectrum
\begin{align}
\Delta \mathbb{E}=
\mathbb{E}(s,0)-\mathbb{E}_{0}=2\psi(3/2)-\psi(2)-\psi(1)%=0.227411\ldots
\end{align}
coincides with the gap in the spectrum of anomalous dimensions
of twist-three light quark-antiquark-gluon operators with large
number of derivatives, see Ref.~\cite{Derkachov:1999ze}.

The $1/N_c^2$ corrections to the ground state energy
$ \mathcal{E}_0(s) = N_c \mathbb{E}_{0} + 1/N_c \delta\mathbb{E}_0$
can be calculated in a standard quantum-mechanical perturbation theory,
evaluating the matrix element $\langle Y^{(0)}|\delta \mathbb{H}|Y^{(0)}\rangle $.
The answer can be written as
\begin{eqnarray}
 \mathcal{E}_0(s) = \mathcal{E}_+(s) + \Delta+O(1/N_c^{3})\,, %\qquad \Delta = N_c - \frac{1}{N_c}\left(3-\frac{\pi^2}3\right)
\label{gap}
\end{eqnarray}
where $\mathcal{E}_+(s)$ is the anomalous dimension for the leading twist DA $\Phi_+$,~Eq.~\eqref{Eplus},
and $\Delta$ is a constant that does not depend on $s$:
\begin{eqnarray} \Delta = N_c + {N^{-1}_c}\left({\pi^2}/3-3\right)\,.
\label{Delta}
\end{eqnarray}
The value of $\Delta$  coincides exactly with the gap between the spectrum of anomalous dimensions of twist-three light quark-antiquark-gluon
operators and the leading-twist quark-antiquark operators for a large number of derivatives, cf.~\cite{Braun:2000av}.

A generic three-particle DA $\Phi_3(z_1, z_2,\mu)$ can be expanded in the eigenfunctions of the large-$N_c$
Hamiltonian
\begin{eqnarray}
\hspace*{-5mm}\Phi_3(z_1, z_2,\mu)
&=&
\int_0^\infty\!\! \!ds \Big[ \eta_0(s,\mu)\,Y^{(0)}_{s}(z_1,z_2)
\nonumber\\&&{}
+ \frac12 \int_{-\infty}^\infty\!\!\!\!\!dx\,\eta(s,x,\mu)\, Y_{s,x}(z_1,z_2)\Big],
\label{3expand}
\end{eqnarray}
where the coefficient functions $\eta_0(s,\mu)$ and $\eta(s,x,\mu)$
can be calculated using the $SL(2)$ scalar product as
\begin{align}
 \eta_0(s,\mu) &= \vev{Y^{(0)}_{s}|\Phi_3}\,,
\notag\\
 \eta(s,x,\mu) &=  x \, \tanh\pi x\, \left(x^2+\frac94\right)\, \vev{Y_{s,x}|\Phi_3}\,.
\label{etas}
\end{align}
They have autonomous scale dependence up to $1/N_c^2$ corrections:
\begin{align}
 \eta_0(s,\mu) &= L^{N_c/\beta_0} R(s;\mu,\mu_0)\,\eta_0(s,\mu_0)\,,
\notag\\
 \eta(s,x,\mu) &=
L^{\gamma_x/\beta_0}R(s;\mu,\mu_0)\,\eta(s,x,\mu_0)\,,
\label{scaledep}
\end{align}
where
$L = {\alpha_s(\mu)}/{\alpha_s(\mu_0)}$,
$\beta_0=\frac{11}3N_c-\frac23 n_f$,
%$\Delta$ is defined in Eq.~(\ref{Delta}),
\begin{align}
  \gamma_x =  N_c\big[\psi\big(3/2+ix\big)+ \psi\big(3/2-ix\big) +2\gamma_E]
\label{gamma}
\end{align}
and
\begin{eqnarray}
R(s;\mu,\mu_0) &=& \exp\left[-\int_{\mu_0}^{\mu} \frac{d\tau}{\tau}\,
\Gamma_{cusp}(\alpha_s(\tau))\,\ln (\tau s/s_0) \right]
\nonumber\\&=& \left(\frac{\mu}{\mu_0}\right)^{-\frac{2C_F}{\beta_0}}
\left(\frac{\mu_0 s}{s_0}\right)^{\frac{2C_F}{\beta_0}\ln L}\!\!
L^{-\frac{4C_F\pi}{\beta_0^2\,\alpha_s(\mu_0)}}
\nonumber\\
\label{Rfactor}
\end{eqnarray}
Here
$s_0= e^{-1/4-\gamma_E}$, $\Gamma_{cusp}(\alpha_s)=\frac{\alpha_s}{\pi} C_F+\ldots $
is the cusp anomalous dimension~\cite{Polyakov:1980ca,Korchemsky:1987wg}, 
and we used that~\cite{Neubert:1993mb}
\begin{align}
 F(\mu) &= L^{-2/\beta_0}  F(\mu_0)\,.  
\end{align} 
Explicit solution of the evolution equation for the DA $\Phi_3(z_1,z_2,\mu)$
in the large-$N_c$ limit presents our main result.

The two-particle DA $\Phi_-(z,\mu)$ can now be recovered from the EOM relation (\ref{EOM}).
Using the expressions for the eigenfunctions in Eqs.~(\ref{Ysx}), (\ref{Y0}) one obtains for the relevant integral:
\begin{eqnarray}
 \lefteqn{ \int_0^z\!\! wdw\,\Phi_3(z, w,\mu)=}
 \\ &=&
-\frac{1}{z^2}\int_0^\infty\!\! \!sds\,e^{is/z}   \int_0^1 du\, u \,
 \biggl[ \eta_0(s,\mu)\,
\nonumber\\&&{}+\dfrac{1}{2}\int_{-\infty}^\infty\!\!\!dx\,\eta(s,x,\mu)\,
{}_2F_1\left(\genfrac{}{}{0pt}{}{-\frac12\!-\!ix,-\frac12\!+\!ix}{2}\Big|-\frac{u}{\bar u}\right)\biggr].
\nonumber
\end{eqnarray}
Remarkably, all terms involving the hypergeometric function vanish thanks to the identity
\begin{align}
 \int_0^1\!\! du\, u\, {}_2F_1\!\left(\genfrac{}{}{0pt}{}{-\frac12\!-\!ix,-\frac12\!+\!ix}{2}\Big|\!-\frac{u}{\bar u}\right)
=0\,,
\end{align}
that is related to the orthogonality condition $\langle Y_{s,x}|Y^{(0)}_{s'}\rangle=\ 0$.
Thus only the ground state (with the lowest anomalous dimension) contributes to the EOM relation (\ref{EOM}).
One finds
\begin{align}
  \int_0^z\! wdw\,\Phi_3(z, w,\mu)
= -\frac{1}{2 z^2}\int_0^\infty \!s ds \,e^{is/z} \,
\eta_0(s,\mu)\,,
\notag
\end{align}
or, equivalently,
\begin{eqnarray}
 \int_0^1 \!du\, u \bar u\, \widetilde \phi_3(s,u,\mu) = -\frac{1}{2s^2} \eta_0(s,\mu)\,,
\end{eqnarray}
leading to the following very simple relation in the $s$-space:
\begin{align}
\widetilde\phi_-(s,\mu)  &=  \widetilde\phi_+(s,\mu) + \eta_0(s,\mu)\,.
\label{tildeEOM2}
\end{align}
Going over from quark/gluon coordinates to the corresponding momenta
\begin{align}
\Phi_\pm(z) &= \int_0^\infty \!d\omega\, e^{-i\omega z} \phi_{\pm}(\omega)\,,
\\
\Phi_3(z_1,z_2) &= \int_0^\infty \!d\omega_1 \,d\omega_2\,e^{-i(\omega_1z_1+\omega_2z_2)} \phi_3(\omega_1,\omega_2)
\notag
\end{align}
can be done easily making use of Eq.~(\ref{Fourier1}).
The $SL(2)$ scalar product in momentum space reads
\begin{align}
\vev{f,g}=2\int\frac{d\omega_1}{\omega_1}\frac{d\omega_2}{\omega_2^2} f^\ast(\omega_1,\omega_2) g (\omega_1,\omega_2)\,,
\end{align}
and the eigenfunctions of the evolution equation take the form
\begin{eqnarray}
 \lefteqn{Y_{s,x}(\omega_1,\omega_2) = }
\notag\\
 &=&-\int_0^1du  \sqrt{ \omega_1}\, J_1\left(2\sqrt{\omega_1 u s}\right)\, \omega_2\,J_2\left(2\sqrt{\omega_2 \bar u s}\right)
 \notag\\
 & &\times \sqrt{s u}\,{}_2F_1\left(\genfrac{}{}{0pt}{}{-\frac12-ix,-\frac12+ix}{2}\Big|-\frac{u}{\bar u}\right)\,.
\end{eqnarray}

In this way one obtains the following expressions for the two-particle DAs in momentum space (cf.~\cite{Braun:2014owa,Feldmann:2014ika})
\begin{align}
\phi_+(\omega,\mu) &=\int_0^\infty ds\,\widetilde\phi_+(s,\mu)\, \sqrt{\omega s}\,J_{1}(2\sqrt{\omega s})\,,
\nonumber\\
\phi_-(\omega,\mu) &=\int_0^\infty ds\, \Big[\widetilde\phi_+(s,\mu)+\eta_0(s,\mu) \Big]\,J_0(2\sqrt{\omega s})\,.
\label{result1}
\end{align}
The scale-dependence of the coefficients $\widetilde\phi_+(s,\mu)$ and $\eta_0(s,\mu)$ differs by a simple overall factor
\begin{align}
\widetilde\phi_+(s,\mu)&= R(s;\mu,\mu_0)\,\widetilde\phi_+(s,\mu_0)\,,
\notag\\
 \eta_0(s,\mu) &= L^{\Delta/\beta_0} R(s;\mu,\mu_0)\,\eta_0(s,\mu_0)\,,
\label{result2}
\end{align}
where $R(s;\mu,\mu_0)$ is defined in Eq.~(\ref{Rfactor}) and $\Delta = N_c + \mathcal{O}(1/N_c)$ is a constant, see Eq.~(\ref{gap}).
In other words, the subleading twist contribution to $\phi_-(\omega,\mu)$ is suppressed at large scales as compared
to the WW contribution by the universal factor $L^{\Delta/\beta_0}$ that does not depend on the light quark momentum.
To the $\mathcal{O}(1/N_c^2)$ accuracy there is no mixing with ``genuine'' quark-gluon degrees of freedom.

It is tempting to define the ``asymptotic'' quark-gluon DA  $\Phi^{\rm as}_3(z_1, z_2,\mu)$ 
as the contribution with the lowest anomalous dimension (for a given $s$):
\begin{eqnarray}
\Phi^{\rm as}_3(z_1, z_2,\mu)
&=& \int_0^\infty\!\! \!ds\, \eta_0(s,\mu)\,Y^{(0)}_{s}(z_1,z_2).
\label{3asy}
\end{eqnarray}
The corresponding expression in momentum space reads
\begin{widetext}
\begin{align}
\phi^{\rm as}_3(\omega_1,\omega_2, \mu) &= -\omega_2\sqrt{\omega_1} \int_0^\infty ds \sqrt{s}\eta^{(0)}(s,\mu)
\int_0^1 du \sqrt{u} J_1\Big(2\sqrt{s\omega_1 u}\Big)J_2\Big(2\sqrt{s\omega_2\bar u}\Big)
\notag\\
&= \frac{\omega_1 \omega_2}{\omega_1+\omega_2}\Big[f_1(\omega_1+\omega_2)-f_0(\omega_1+\omega_2)\Big]
+\omega_1\Big[f_1(\omega_1+\omega_2)-f_1(\omega_1)\Big],
\label{phi3-as}
\end{align}
\end{widetext}
where
\begin{align}
f_k(\omega)&=\int_0^\infty ds\,\eta^{(0)}(s,\mu)\big(\sqrt{\omega s}\big)^{-k}J_k\big(2\sqrt{\omega s}\big)\,.
\end{align}

\section{Asymptotic behavior at small and large momenta}

One of the reasons why the renormalization group evolution is interesting is that
it gives insight in the expected behavior of the DAs at large and small momenta,
which is important for the status of factorization theorems.
Although one cannot make any rigorous statements on the shape of the DAs at low scales,
it is usually assumed that the ``true'' DAs have the same asymptotic behavior as
in perturbation theory. This assumption proved to be successful for modeling of parton distributions
and DAs of light hadrons, so that it is natural to use the same logic for heavy-light systems.

For small momenta there are no surprises. Using explicit expressions we find
\begin{align}
 \phi^{\rm as}_3(\omega_1,\omega_2, \mu) \stackrel{\omega_1\to 0}{\sim} \mathcal{O}(\omega_1)\,,
\notag\\
 \phi^{\rm as}_3(\omega_1,\omega_2, \mu) \stackrel{\omega_2\to 0}{\sim} \mathcal{O}(\omega_2^2)\,,
\end{align}
respectively.
This behavior is in agreement with arguments based on quark-gluon duality~\cite{Khodjamirian:2006st}.
If both quark and gluon momenta are small one obtains
$$
\phi^{\rm as}_3(\omega_1,\omega_2, \mu) \stackrel{\omega_1,\omega_2\to 0}{=}
 -\frac{\omega_1 \omega_2^2}{12}\int_0^\infty ds\,s^2 \, \eta^{(0)}(s,\mu)\,.
$$
The large-momentum asymptotics is much more interesting.
An inspection of the the asymptotic DA (\ref{phi3-as}) reveals that it does not decrease for large gluon momenta
$\omega_2\to\infty$ (because of the last term that is $\omega_2$-independent).
As a consequence, integral over all momenta is ill-defined, and the normalization of the asymptotic DA to a
matrix element of a local operator even at a single scale is not possible.
This problem is seen even better in coordinate space.  Using the definition in (\ref{3asy}) and
explicit expression for $Y^{(0)}_{s}(z_1,z_2)$ one obtains
\begin{eqnarray}
\Phi^{\rm as}_3(z_1, z_2)
&=& \frac{1}{z_2}\partial_{z_1}\partial_{z_2}\frac{z_1 z_2}{z_1\!-\!z_2}\Big[\Xi(z_1)-\Xi(z_2)\Big],
\label{3asy2}
\end{eqnarray}
where
\begin{align}
\Xi(z)=\int_0^\infty \frac{ds}{s} \,\eta_0(s) e^{is/z}
\end{align}
The behavior of $\Xi(z)$ at $z\to 0$ is determined by the small-$s$ asymptotics of $\eta_0(s)$.
Assuming a power-law behavior $\eta_0(s)\stackrel{s\to 0}{\sim} s^a$ one obtains
\begin{align}
\Phi_3(z_1,z_2)  \sim \frac{1}{z_2}\partial_{z_1}\partial_{z_2}z_1 z_2 \frac{z_1^a- z_2^a}{z_1-z_2}\,.
\end{align}
This function is not analytic at the origin $\{z_1,z_2\}=0$: the limit $z_i\to 0$ depends on the way
the variables approach zero and exists only if $a\geq 2$. For $a=2$ one gets
\begin{align}
\Phi_3(z_1,z_2)\sim 2\big[1+z_1/z_2\big] + \mathcal{O}\big((z_1/z_2)^2, z_2\big)\,.
\end{align}
If the gluon coordinate $z_2\to 0$ and the quark position $z_1$ is kept constant, the singularity $\sim 1/z_2$
cannot be avoided. It translates to the constant behavior at large gluon momentum, as seen explicitly from the momentum space
representation.

The singular behavior $\sim 1/z_2$, corresponding in physics terms to the instability due to 
gluon falling to the center of the color-Coulomb field,
is not a special pathology of the asymptotic DA: the contributions of the continuum spectrum are even
more singular, $\sim (1/z_2)^{3/2\pm ix}$, so that the corresponding momentum space DAs are increasing (and oscillating)
functions of the gluon momentum. We are able to show that all such singularities are,  however, spurious and cancel in the
sum of contributions of the asymptotic DA and the corrections. Most importantly, this cancellation is
not spoiled by the evolution: The $\sim 1/z_2$ singularity is not generated at higher scales provided it is not
present already in the nonperturbative ansatz at a reference low scale. This result implies that for small $z_2$, alias
large $\omega_2 \gtrsim \mu $, the hierarchy of contributions with increasing anomalous dimensions is lost; the leading
large-$\omega_2$ asymptotics of the ``asymptotic'' DA is exactly cancelled by the contributions with larger anomalous dimensions.
This pattern appears to be unconventional, we are not aware of examples of similar behavior for light quark systems.

To this end we rewrite the expansion (\ref{3expand}) as the integral over imaginary axis $x \mapsto -i x$
\begin{widetext}
 \begin{eqnarray}
\Phi_3(z_1,z_2) & = & \int_0^\infty ds\, \biggl[Y^{(0)}_s(z_1,z_2)\, \vev{Y^{(0)}_s|\Phi_3}
+ \frac{1}{2i}\int_{-i\infty}^{i\infty} dx \, x\,
     \tan\pi x \,\left(x^2-\frac94\right)\mathbb{Y}_{s,x}(z_1,z_2)\vev{\mathbb{Y}_{s,x}|\Phi_3} \biggr]
\notag\\
& = &  \frac{1}{2i} \int_0^\infty ds \int_{-1 -i\infty}^{-1 + i\infty} dx \, x\,
     \tan\pi x \,\left(x^2-\frac94\right)\mathbb{Y}_{s,x}(z_1,z_2)\vev{\mathbb{Y}_{s,x}|\Phi_3}\,,
\label{imaginary}
\end{eqnarray}
\end{widetext}
where
\begin{align}
\mathbb{Y}_{s,x}(z_1,z_2)= Y_{s,-ix}(z_1,z_2)\,.
%=\frac{is^2}{z_1^2 z_2^3}\int_0^1 du\,u\bar  u^{\frac12-x}\, e^{is(u/z_1+\bar u/z_2)}
%\frac{\Gamma(2x)}{\Gamma(-\frac12+x)\Gamma(\frac52+x)} {}_2F_1\left(\genfrac{}{}{0pt}{}{-\frac12-x,\frac52-x}{1-2x}\Big|{\bar u}\right)
\end{align}
Note that the contribution of the asymptotic DA is taken into account in the second
line of Eq.~(\ref{imaginary}) by moving the integration contour to the left of the singularity at $x=1/2$,
due to $\tan \pi x$. This representation remains valid after
the scale dependence (\ref{scaledep}) of the coefficients is
taken into account: The anomalous dimension $\gamma_{-ix}$ (\ref{gamma}) is an analytic function in the strip
$ -3/2<\text{Re}\, x<3/2$ and $\gamma(\pm i/2)$ gives the anomalous dimension of the asymptotic DA.

In order to study the limit $z_2\to 0$ we write
\begin{align}
\mathbb{Y}_{s,x}(z_1,z_2) &= \Upsilon_{s,x}(z_1,z_2)+\Upsilon_{s,-x}(z_1,z_2)
\end{align}
where
\begin{widetext}
\begin{eqnarray}%
\Upsilon_{s,x}(z_1,z_2)&=&\frac{is^2}{z_1^2 z_2^3}\int_0^1 du\,u\bar  u^{\frac12-x}\, e^{is(u/z_1+\bar u/z_2)}
\frac{\Gamma(2x)}{\Gamma(x-\frac12)\Gamma(x+\frac52)}
 {}_2F_1\left(\genfrac{}{}{0pt}{}{-\frac12-x,\frac52-x}{1-2x}\Big|{\bar u}\right)
\end{eqnarray}
and use that the coefficient $\vev{\mathbb{Y}_{s,x}|\Phi_3}$ in (\ref{imaginary}) is symmetric under $x\to -x$ so
that one can replace
$\mathbb{Y}_{s,x}(z_1,z_2) \to 2\Upsilon_{s,x}(z_1,z_2)$ without changing the value of the integral.
For small $z_2$
\begin{align}
\Upsilon_{s,x}(z_1,z_2)\sim\frac{i}{z_1^2 z_2} \left(\frac s{z_2}\right)^{1/2+x} e^{is/z_1}
\frac{\Gamma(2x)\Gamma(3/2-x)}{\Gamma(-\frac12+x)\Gamma(\frac52+x)}\Big[1+\mathcal{O}(z_2)\Big],
\end{align}
\end{widetext}
so that the asymptotic behavior of the DA $\Phi_3(z_1,z_2)$ at $z_2\to 0$ is determined by the position of the
closest to the origin singularity of the integrand  on the real negative $x$ axis, to the
left of the integration contour at $\text{Re}\, x =  - 1 $.
In this way the pole at  $ x =  - 1/2 $ corresponding the $\sim 1/z_2$ behavior
is always avoided and the closest singularity appears to be at $x =  - 3/2 $, corresponding to
$\Phi_3(z_1,z_2\to 0) \sim  \text{const}$, unless the matrix element
$\vev{\mathbb{Y}_{s,x}|\Phi_3}$ is more singular.

Taking into account the scale dependence amounts to the insertion of the RG factors (\ref{scaledep}) under the
integral. In this way additional singularities appear corresponding to the poles of the anomalous
dimension~(\ref{gamma}),
$\gamma_{-ix} = N_c[\psi(3/2+x)+\psi(3/2-x)+2\gamma_E]$. The singularity closest to the origin is at $x=-3/2$
so that if the initial condition for the evolution  $\Phi_3(z_1,z_2,\mu_0)$ corresponds to a constant 
behavior at $z_2\to 0$, it will be modified to $\Phi_3(z_1,z_2\to 0 ) \sim  \ln (\mu z_2)$, 
corresponding to a ``tail'' $1/\omega_2$   in momentum space. The same behavior was found previously for the leading twist DA
~\cite{Lange:2003ff,Braun:2003wx,Lee:2005gza}. It is easy to see that in the other limit
$\Phi_3(z_1\to 0,z_2) \sim  \ln (\mu z_1)$ as well, so that our final conclusion is that gluon emission generates a
radiative tail  $\sim 1/\omega_1$ and/or $\sim 1/\omega_2$ of the three-particle DA $\phi_3(\omega_1,\omega_2,\mu)$
for both, large light quark and large gluon momenta. This is natural as the corresponding terms are present in the evolution kernels.
The reason and consequences of such a behavior have been discussed at length in the literature, e.g. \cite{Braun:2003wx}, so we
do not need to repeat this discussion here.

\section{A simple model}

For the simplest phenomenologically acceptable model of the leading-twist B-meson DA at a low scale
$\mu=\mu_0$ one usually takes~\cite{Grozin:1996pq}
\begin{align}
 \phi_+(\omega) = \frac{\omega}{\lambda_B^2}e^{-\omega/\lambda_B}
&&\mapsto &&
\widetilde\phi_+(s) = e^{-s\lambda_B}\,,
\label{model+}
\end{align}
where $\lambda_B$ is defined as
\begin{align}
  \frac{1}{\lambda_B} & = \int_0^\infty\!\frac{d\omega}{\omega}\,\phi_+(\omega)\,.
\end{align}
The value of $\lambda_B$ is the most important nonperturbative parameter in the QCD factorization approach~\cite{Beneke:1999br,Beneke:2000wa},
with current estimates in the range $\lambda_B \simeq 300-600$~MeV~\cite{Braun:2003wx,Nishikawa:2011qk}.

In the same spirit, we consider a simple  model for the three-particle DA at a reference scale
\begin{align}
 \phi_3(\omega_1,\omega_2,\mu_0) &= \frac{\varphi_3}{\omega_0^5}\, \omega_1 \omega_2^2 \, e^{-(\omega_1+\omega_2)/\omega_0}\,,
\notag\\ \mapsto \quad
\widetilde\phi_3(s,u,\mu_0)   &= \varphi_3 e^{-s\omega_0}\,,
\label{model3}
\end{align}
where $\varphi_3$ is a constant that can be related to the matrix elements of local quark-gluon operators~\cite{Grozin:1996pq,Nishikawa:2011qk}
\begin{align}
\varphi_3 = \frac16 \Big[\lambda^2_E-\lambda^2_H\Big] \,.
\end{align}
The recent QCD sum rule calculation~\cite{Nishikawa:2011qk} gives $\lambda^2_E-\lambda^2_H = -0.03\pm 0.03$~GeV$^2$.
The corresponding DA in
coordinate space is
\begin{align}
 \Phi_3(z_1,z_2,\mu_0) = \frac{2\varphi_3}{(1+i \omega_0 z_1)^2(1+i\omega_0 z_2)^3}
\end{align}
and the two-particle DA $\phi_-(\omega,\mu_0)$ for this model can be obtained directly from the EOM:
\begin{eqnarray}
\lefteqn{\phi_-(\omega,\mu_0) =}
\nonumber\\&=& \frac{e^{-\omega/\lambda_B}}{\lambda_B} - \frac23\frac{\varphi_3}{\omega_0^3} e^{-\omega/\omega_0}
\Big[1-2\frac{\omega}{\omega_0}+\frac12\frac{\omega^2}{\omega_0^2}\Big],
\end{eqnarray}
where the first term is the WW contribution related to the leading-twist DA.
The higher-twist contribution $\sim \varphi_3$ in the $s$-space reads
\begin{align}
 \eta_0(s,\mu_0) &= - \frac{1}3 \varphi_3 s^2 e^{-\omega_0 s},
\end{align}
and the DA at higher scales can easily be calculated from Eq.~(\ref{result1}), (\ref{result2}).
The higher-twist contribution is suppressed at large scales by an overall factor $(\alpha_s(\mu)/\alpha_s(\mu_0))^{\Delta/\beta_0}$ (\ref{scaledep})
as compared to the leading twist.

Let us now have a closer look at the three-particle DA (\ref{model3}) itself.
The asymptotic DA corresponding to this model is
\begin{eqnarray}
 \phi^{\rm as}_3(\omega_1,\omega_2) &=&
\frac{\phi_3 \omega_1}{3\omega_0^4} e^{-\frac{\omega_1+\omega_2}{\omega_0}}
  \Big[
\big(\omega_1-2\omega_0\big) \left(1-e^{\frac{\omega_2}{\omega_0}}\right)
\notag\\
&&{}
+\frac{\omega_2}{\omega_0}\big(\omega_2+\omega_1-2\omega_0\big)\Big].
\end{eqnarray}
The shape of $\phi^{\rm as}_3(\omega_1,\omega_2)$ is qualitatively different from
$\phi_3(\omega_1,\omega_2)$: it is not factorizable as a product of the distributions depending on
$\omega_1$ and $\omega_2$, does not decrease at $\omega_2\to \infty$ and becomes negative for large $\omega_1$.
This is illustrated in Fig.~1 where we show the ratio $\phi^{\rm as}_3/\phi_3$ as a function of
$\omega_2/\omega_0$ for several different values of $\omega_1/\omega_0$.

\begin{figure}[t]
\centerline{
\epsfig{file=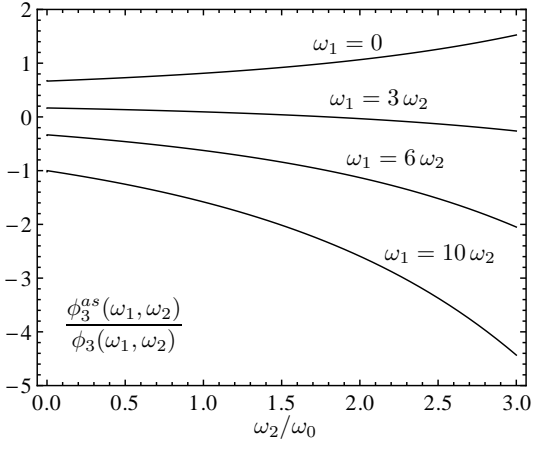,scale=0.9}
%\begin{picture}(210,175)(0,0)
%\put(-5,0){\epsfxsize7.8cm\epsffile{Rphi3.eps}}
%\put(100,-8){$\omega_2/\omega_0$}
%\put(20,35){$\dfrac{\phi_3^{as}(\omega_1,\omega_2)}{\phi_3(\omega_1,\omega_2)}$}
%\put(125,153){$\omega_1 = 0$}
%\put(132,130){$\omega_1 = 3\, \omega_2$}
%\put(139,105){$\omega_1 = 6\, \omega_2$}
%\put(155,65){$\omega_1 = 10\, \omega_2$}
%\end{picture}
}
\caption{ The ratio $\phi^{\rm as}_3/\phi_3$ as a function of $\omega_2/\omega_0$ for  several values of
$\omega_1/\omega_0$ for the model in Eq.~(\ref{model3}).
}
\label{fig:MRmodel}
\end{figure}

\begin{figure}[t]
\centerline{
\epsfig{file=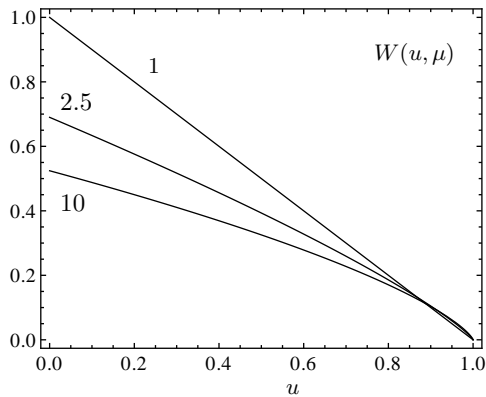,scale=0.95}
%\begin{picture}(210,140)(0,0)
%\put(-5,0){\epsfxsize0.4\textwidth\epsffile{Phiu.eps}}
%\put(105,-8){$  u $}
%\put(140,125){$W(u,\mu)$}
%\put(50,120){$1$}
%\put(15,105){$2.5$}
%\put(15,65){$10$}
%\end{picture}
}
\caption{The function $W(u,\mu)$, Eq.~\eqref{Phiu}, on three different scales 
$\mu=\mu_0=1\,\text{GeV}$, $\mu = 2.5\,\text{GeV}$ and  $\mu=10\,\text{GeV}$.
}
\label{fig:Phiu}
\end{figure}

The quark-gluon DA at higher scales is given by
\begin{widetext}
 \begin{align}
      \phi_3(\omega_1,\omega_2,\mu) & =
     \frac{1}{2i} \int\limits_0^\infty ds\, R(s;\mu,\mu_0) \int_{C} %\limits_{-1 -i\infty}^{-1 + i\infty}
     dx \, x\,
     \tan\pi x \,\left(x^2-\frac94\right)\mathbb{Y}_{s,x}(\omega_1,\omega_2)
%\nonumber\\&&{} \times
L^{\gamma_{-ix}/\beta_0}
      \vev{\mathbb{Y}_{s,x}|\Phi_3}\,,
\label{imaginary3}
\end{align}
where the $x-$integration goes along the imaginary axis with $\text{Re}(x)=-1$.
For numerical evaluation it is convenient to consider first the corresponding function in the $(s,u)$-representation~\eqref{tildePsi},
\begin{align}
 \bar u\,\widetilde \phi_3(s,u,\mu)= -\frac{1}{2i s^2}  R(s;\mu,\mu_0) \int_C%\limits_{-1 -i\infty}^{-1 + i\infty}
dx \, x\,
     \tan\pi x \,\left(x^2-\frac94\right)\,
{}_2F_1\left(\genfrac{}{}{0pt}{}{-\frac12-x,-\frac12+x}{2}\Big|-\frac{ u}{\bar u}\right)L^{\gamma_{-ix}/\beta_0}
      \vev{\mathbb{Y}_{s,x}|\Phi_3}\,,
\end{align}
and calculate the DA in momentum space as 
\begin{align}
\phi(\omega_1,\omega_2)=\int_0^\infty ds \,s \int_0^1 du \,(\omega_1 s u)^{1/2}\, ( \omega_2 s \bar u)\, J_1\left(2\sqrt{us \omega_1}\right)
J_2\left(2\sqrt{\bar us \omega_2}\right)\,\widetilde \phi_3(s,u,\mu)\,.
\end{align}
\end{widetext}
For our model (\ref{model3})
\begin{align}
\vev{\mathbb{Y}_{s,x}|\Phi_3} &= \frac{1}3 \varphi_3 s^2 e^{-\omega_0 s}\frac\pi{\cos\pi x}\left(x^2-\frac14\right)\,
\end{align}
and the function $\widetilde \phi_3(s,u,\mu)$ has a factorized form
\begin{align}
 \bar u\,\widetilde \phi_3(s,u,\mu)=\varphi_3 R(s;\mu,\mu_0) e^{-\omega_0 s} W(u,\mu)\,,
\end{align}
where
\begin{align}\label{Phiu}
W(u,\mu) &=\frac{i}{6}
\int_C%\limits_{-1 -i\infty}^{-1 + i\infty}
dx \, \frac{\pi x \sin\pi x}{\cos^2\pi x}\,
      \,\left(x^2-\frac94\right)\left(x^2-\frac14\right)
     \notag\\
&\times{}_2F_1\left(\genfrac{}{}{0pt}{}{-\frac{1}{2}-x,-\frac{1}{2}+x}{2}\Big|-\frac u{\bar u}\right)L^{\gamma_{-ix}/\beta_0}.
\end{align}
The function $W(u,\mu)\propto  \bar u\,\widetilde \phi_3(s,u,\mu) $ at three different scales 
$\mu=\mu_0=1\,\text{GeV}$, $\mu = 2.5\,\text{GeV}$ and  $\mu=10\,\text{GeV}$ is shown in Fig.~\ref{fig:Phiu}.
Note that $u$ and $\bar u = 1-u$ have the meaning of the momentum fractions
carried by the quark and the gluon, respectively:
\begin{align}
 \omega_1 = u\omega\,, &&  \omega_2 = \bar u\omega\,, && \omega = \omega_1+\omega_2\,,
\label{omega-u}
\end{align}  
so that the scale dependence visualized in Fig.~\ref{fig:Phiu} corresponds to a redistribution 
of the total momentum of the light degrees of freedom such that at large scales the gluon carries
a larger fraction. Note also that  $W(u,\mu)$ becomes slightly curved at $u\to 1$ but still vanishes,
or, equivalently, the DA $\widetilde \phi_3(s,u,\mu)$ diverges in the same limit, but the divergence is
softer than a power $1/\bar u$.    

\begin{figure*}[ht]
%\begin{picture}(180,140)(0,0)
%\put(-20,0){\epsfxsize0.42\textwidth\epsffile{omega0p3.eps}}
%\put(85,-8){$  u $}
%\put(125,100){$\widehat\phi_3(u,\omega,\mu)$}
%\put(55,30){$\omega=0.3\,\omega_0$}
%\end{picture}
\epsfig{file=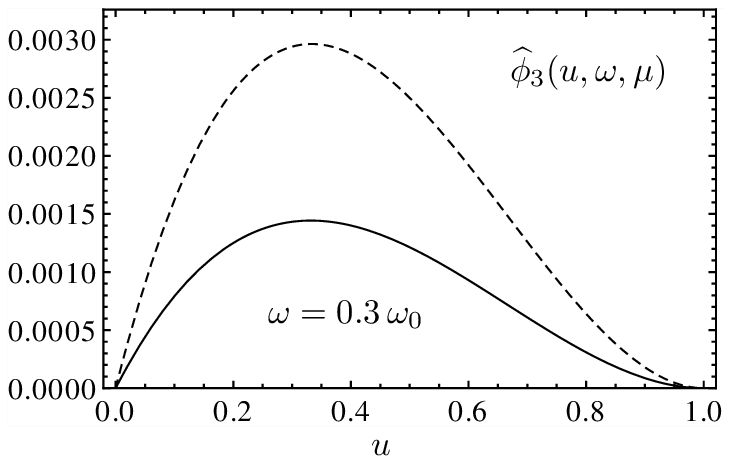, scale=1}\hspace*{0.32cm}
\epsfig{file=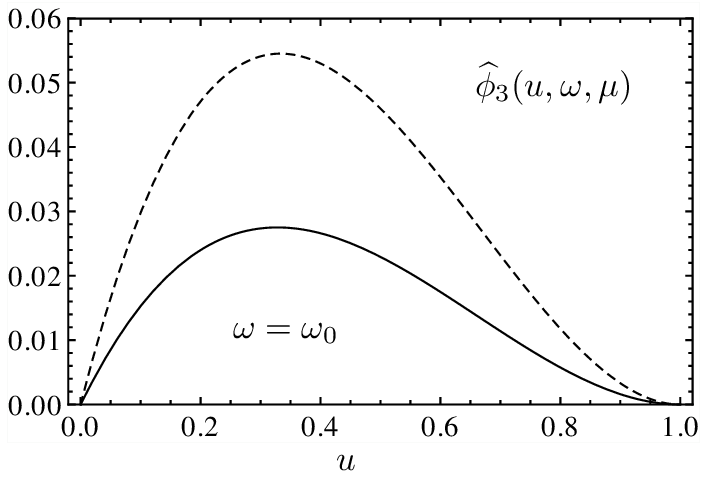, scale=1}\\
%\hspace*{0.5cm}
%\begin{picture}(180,140)(0,0)
%\put(10,0){\epsfxsize0.40\textwidth\epsffile{omega1.eps}}
%\put(105,-8){$  u $}
%\put(145,100){$\widehat\phi_3(u,\omega,\mu)$}
%\put(75,30){$\omega=\omega_0$}
%\end{picture}
%\\

%\hspace*{0.2cm}
\hspace*{0.38cm}\epsfig{file=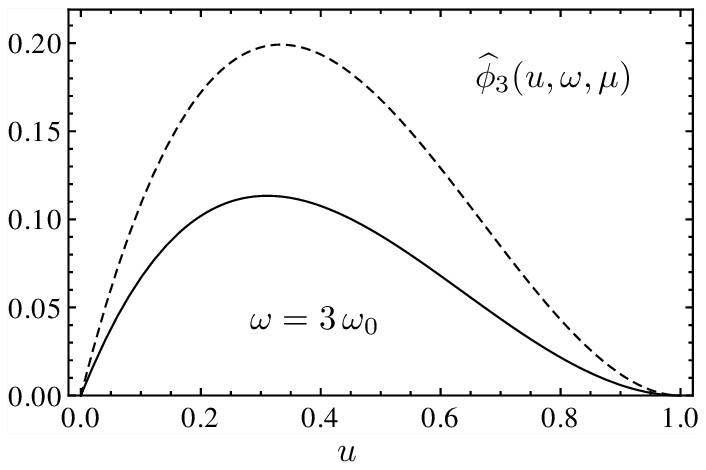,scale=1}
%\hspace*{0.1cm}
\epsfig{file=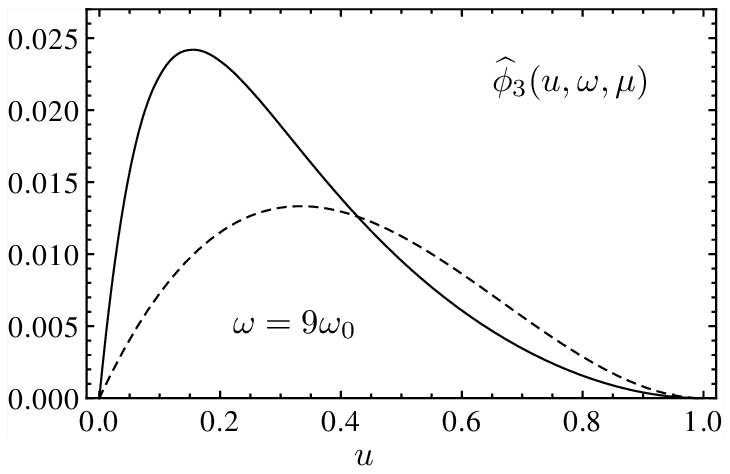,scale=1}
%\begin{picture}(180,140)(0,0)
%\put(-20,0){\epsfxsize0.40\textwidth\epsffile{omega3.eps}}
%\put(75,-8){$  u $}
%\put(115,100){$\widehat\phi_3(u,\omega,\mu)$}
%\put(50,30){$\omega= 3\,\omega_0$}
%\end{picture}
%
%\begin{picture}(180,140)(0,0)
%\put(10,0){\epsfxsize0.41\textwidth\epsffile{omega9.eps}}
%\put(110,-8){$  u $}
%\put(150,100){$\widehat\phi_3(u,\omega,\mu)$}
%\put(75,30){$\omega=9\omega_0$}
%\end{picture}
\caption{The three-particle B-meson DA (\ref{hatphi3}) as a function of the quark momentum fraction $u$ 
 after the evolution to $\mu =2.5$~GeV (solid curves) and at the 
initial scale $\mu=\mu_0=1$~GeV (dashed curves) for the model in Eq.~(\ref{model3}) for several values 
of the total momentum $\omega=\omega_1+\omega_2$. 
}
\label{fig:evol}
\end{figure*}

Finally, we show in  Fig.~\ref{fig:evol} the (normalized) DA in momentum space 
\begin{align}
 \widehat\phi_3(u,\omega,\mu) = \frac{\omega_0^2}{\varphi_3} \phi_3(u\omega, \bar u\omega,\mu)
\label{hatphi3}
\end{align}
as a function of the quark momentum fraction after the evolution to $\mu =2.5$~GeV (solid curves) and at the 
initial scale $\mu=\mu_0=1$~GeV (dashed curves) 
for four different values of the total momentum $\omega/\omega_0 = \{0.3, 1, 3, 9\}$.

One sees that for small total momentum the DA is rather strongly suppressed by the evolution whereas the shape is only weakly affected.
On the contrary, for large momentum there is no suppression and the DA is strongly tilted towards small values of $u$, corresponding 
to small quark and large gluon momenta. 
It would be interesting to analyze the large-momentum behavior using the expansion of the type suggested in~\cite{Lee:2005gza} 
(see also~\cite{Feldmann:2014ika}). Such a study goes, however, beyond the tasks of this paper.

\vspace*{5mm}

\section{Conclusions}

To summarize, we have shown that
the evolution equation for the three-particle quark-gluon B-meson light-cone DA of subleading twist is completely integrable 
in the large $N_c$ limit and can be solved exactly in analytic form. The most important result for phenomenology is that 
 ``genuine'' three-particle contributions of quark-gluon states essentially decouple
from the subleading-twist two-particle DA $\phi_-(\omega)$~\cite{Beneke:2000wa} so that its properties are similar to the
leading-twist DA. A similar simplification has been found before for the structure function $g_2(x,Q^2)$ in polarized 
deep-inelastic lepton-hadron scattering~\cite{Ali:1991em,Balitsky:1996uh,Braun:2000av,Braun:2001qx}.
Based on this experience, we expect that ``genuine'' three-particle contributions do not contribute directly to 
many physical observables in B-decays at tree level
because three-particle and two-particle twist-three contributions to the products of currents are typically related by Ward identities;
hence they cannot have a different scale dependence. 
We also expect that a similar simplification of the renormalization-group dependence holds for twist-four distributions as well. 
This study is in progress \cite{BM}. In this way  we hope to be able to identify important degrees of freedom 
in multiparticle quark-gluon distributions in heavy mesons that can be parametrized by a minimum number of 
nonperturbative parameters. This would present a step forward towards understanding of subleading corrections
in powers of the heavy quark mass and ultimately allow one to increase significantly 
the accuracy of QCD predictions for heavy meson (and baryon) decays based on the heavy-quark expansion 
and/or light-cone sum rules.

%%%%%%%%%%%%%%%%%%%%%%%%%%%%%%%%%%%%%%%%%%%%%%%%%%

\vskip5mm

%{\large \bf 7.}~~
%{\large\bf Acknowledgements}\\
\section*{Acknowledgements}
The work of A.M. was supported by the DFG, grant MO~1801/1-1.

\end{document}